\documentclass[twocolumn,prb,showpacs,preprintnumbers,amsmath,amssymb]{revtex4}
\usepackage{graphicx}
\usepackage{dcolumn}
\usepackage{bm}

\begin{document}



\title{Temperature and magnetization dependent band-gap
  renormalization and optical many-body effects in diluted magnetic
  semiconductors}

\author{Ying Zhang} 

\author{S. Das Sarma} 

\affiliation{Condensed Matter Theory Center, Department of Physics,
  University of Maryland, College Park, MD 20742-4111}

\date{\today}

\begin{abstract}
We calculate the Coulomb interaction induced density, temperature and
magnetization dependent many-body band-gap renormalization in a
typical diluted magnetic semiconductor ${\rm Ga}_{1-x}{\rm Mn}_x{\rm
  As}$ in the optimally-doped metallic regime as a function of carrier
density and temperature. We find a large ($\sim 0.1 eV$) band gap
renormalization which is enhanced by the ferromagnetic transition.  We
also calculate the impurity scattering effect on the gap narrowing. We
suggest that the temperature, magnetization, and density dependent
band gap renormalization could be used as an experimental probe to
determine the valence band or the impurity band nature of carrier
ferromagnetism.
\end{abstract}


\maketitle

\section{Introduction}

A central open question~\cite{DMS} in the physics of carrier-mediated
ferromagnetism in diluted magnetic semiconductors (DMS) is the nature
of the semiconductor carriers mediating the ferromagnetic interaction,
in particular, whether the carriers (i.e. the holes in the
well-studied Ga$_{1-x}$Mn$_x$As system) are itinerant band carriers
(i.e. holes in the GaAs valence band for Ga$_{1-x}$Mn$_x$As) residing
in the (valence) band of the parent semiconductor or impurity band
carriers residing in a narrow impurity band within the band gap of the
semiconductor. This issue is of fundamental
significance~\cite{chattopadhyay} since the valence band~\cite{dietl}
or impurity band~\cite{berciu} competing pictures respectively imply
competing RKKY-Zener~\cite{dietl2} or double exchange~\cite{akai}
mechanisms for DMS ferromagnetism.  In spite of intense experimental
and theoretical activity in Ga$_{1-x}$Mn$_x$As over the last ten
years, the key issue of valence band versus impurity band carriers
mediating the DMS ferromagnetism has remained controversial even in
the optimally doped, ($x \sim 0.05$) nominally metallic, high-$T_c$
($T_c \sim 150 - 200K$, the ferromagnetic transition temperature or
the Curie temperature) Ga$_{1-x}$Mn$_x$As material. For example, {\it
  ab initio} first principles band structure
calculations~\cite{sanvito} typically indicate a strong-coupling
narrow impurity band behavior whereas the extensively used
phenomenological mean-field description, parameterized by a single
effective impurity moment-carrier spin (``pd'') exchange
interaction~\cite{DMS, dietl, berciu, dassarma2}, leads to reasonable
quantitative agreement with experimental results in the metallic ($x
\sim 0.05$) Ga$_{1-x}$Mn$_x$As requiring a relatively weak exchange
coupling (and therefore, a weak perturbation of the GaAs valence band)
between the Mn moments and the valence band holes. Similarly, optical
absorption spectroscopic data in GaMnAs were first
interpreted~\cite{hirakawa} using an impurity band theoretic
description~\cite{hwang}, but later it was shown~\cite{yang} that the
same data could also be explained as arising from the valence band
picture. Given the great complexity of many competing interaction and
disorder effects in the DMS Hamiltonian, it is increasingly clear that
this important question cannot be settled purely by theoretical means,
and an unambiguous experimental test is warranted.

In this paper we propose an experimental (optical) measurement of the
carrier-induced many-body band gap renormalization (BGR) in
Ga$_{1-x}$Mn$_x$As for the definitive resolution of this controversy;
in particular, we establish theoretically that the BGR in
Ga$_{1-x}$Mn$_x$As should be {\em extremely} large ($\sim 0.1 eV$) and
a strong function of hole density if the carriers are indeed GaAs
valence band holes, allowing for a clear distinction between the
(valence band) RKKY-Zener and the (impurity band) double exchange
mechanisms for DMS ferromagnetism. In addition, we calculate the
temperature (as well as density) and hole spin-polarization dependent
BGR in Ga$_{1-x}$Mn$_x$As, finding a strong quantitative dependence of
the BGR on the magnetic properties -- in particular, the calculated
BGR depends strongly on whether the system is ferromagnetic or
paramagnetic with the ferromagnetic Ga$_{1-x}$Mn$_x$As typically
having a factor of $1.5$ to $2$ higher BGR than the corresponding
paramagnetic system at the same density (and temperature). Our
predicted density, temperature, and spin-polarization (i.e.
ferromagnetic being fully spin-polarized and paramagnetic being
spin-unpolarized) dependence of BGR in Ga$_{1-x}$Mn$_x$As (provided
the carriers are the usual GaAs valence band holes) should enable a
clear distinction between the valence band and the impurity band
picture of DMS ferromagnetism through a careful analysis of
experimental optical absorption data~\cite{singley}. We want to
emphasize that our work is not only about studying BGR in the
Ga$_{1-x}$Mn$_x$As system. Our main purpose is to propose a technique
for distinguishing whether the GaMnAs carriers are in the valence band
or the impurity band, which is a subject of considerable importance
and controversy in the DMS community as exemplified by many published
works on the topic. On the other hand, to our knowledge the
temperature and magnetization dependence of the BGR are never studied
before, and they are crucial to the experimental observation of BGR.

Free carriers (e.g. holes in Ga$_{1-x}$Mn$_x$As) affect the band gap
of semiconductors in two essential ways. First, there is the trivial
single-particle effect of band filling (sometimes called the
Moss-Burstein shift) arising from a finite Fermi level $E_F$ ($\propto
n^{2/3}$, where $n$ is the free carrier density) in the valence or the
conduction band of the semiconductor as the band is filled with a
finite carrier density. The band filling effect obviously increases
the apparent band gap by an amount $E_F$ which should be subtracted
out from the measured band gap energy. In our results presented in
this paper we subtract out the trivial band filling Moss-Burstein
shift in order to avoid any confusion. The second effect, which is the
main subject of this work, is the self-energy correction of the
semiconductor band edge due to the many-body exchange-correlation
effect arising from finite carrier density. The band gap
renormalization~\cite{dassarma3, trankle} is a true many-body effect
arising from the hole-hole Coulomb interaction in the
Ga$_{1-x}$Mn$_x$As valence band, with the intrinsic (i.e. the zero
carrier density limit) band gap being increasingly reduced (i.e.
renormalized) by the carrier-induced self-energy correction as the
carrier density increases. This density dependent many-body
self-energy correction induced reduction or renormalization of the
semiconductor band gap, the BGR, in Ga$_{1-x}$Mn$_x$As is the subject
matter of our work.

The rest of the paper is structured as following: In Section
\ref{sec:form} we present the formalism of our BGR calculation as a
function of temperature, carrier density and spin degeneracy. In
Section \ref{sec:results} we describe the result of BGR for an
optimally-doped metallic Ga$_{1-x}$Mn$_x$As DMS system. In Section
\ref{sec:diss} we discuss the implication of our BGR results how they
might serve as a probe to determine the valence band or impurity band
nature of carrier ferromagnetism, which is the purpose of this paper.
Especially, we discuss the effect of impurity scattering on the
Ga$_{1-x}$Mn$_x$As optical properties, and why in spite of this
complication, the many-body BGR effect can still be experimentally
observed.

\section{Formalism}
\label{sec:form}

\begin{figure}[htbp]
\centering \includegraphics[width=2in]{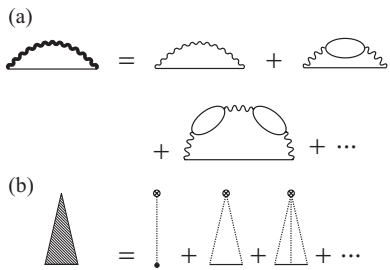}
  \caption{Feynman diagrams for calculating the quasihole
    self-energy. The solid line denotes the hole propagator. (a)
    one-loop quasihole self-energy due to Coulomb interaction, with
    the wiggly line denoting long-range Coulomb interaction; (b)
    quasihole self-energy due to impurity scattering in the
    single-site approximation, with the dashed line denoting
    short-range impurity interaction.}
  \label{fig:feynman}
\end{figure}

We calculate the BGR in the single-loop (Fig.~\ref{fig:feynman})
self-energy theory which is the leading-order approximation in the
dynamically screened Coulomb interaction. This approximation, often
referred to as the ``GW approximation''~\cite{hedin} in the
semiconductor literature~\cite{louie}, is known to be exact in the
high carrier density limit which happens to be the situation of
interest to us since Ga$_{1-x}$Mn$_x$As is an extremely heavily
hole-doped system with the hole density $n \approx 10^{20} cm^{-3}$.
In Ga$_{1-x}$Mn$_x$As the free carriers are holes, so we only consider
the self-energy correction to the band gap arising from the quasihole
self-energy correction to the valence band edge within the
GW-approximation (Fig.~\ref{fig:feynman}), obtaining for the
finite-temperature Matsubara self-energy ($\hbar = k_B = 1$
throughout):

\begin{eqnarray}
  \label{eq:self}
  \Sigma({\bf k}, i \nu_l) = - \int \frac{d^3 q}{(2 \pi)^3}
    T \sum_{\omega_n} \frac{v_q}{\epsilon({\bf q}, i \omega_n)}
      \nonumber \\
      \times \frac{1}{i \nu_l + i \omega_n - \xi_{{\bf q} - {\bf k}}},
\end{eqnarray}
where $v_q = 4 \pi e^2 / \kappa q^2$ is the 3D bare Coulomb
interaction with $\kappa$ the lattice dielectric constant, $i \nu_l =
i (2l + 1) \pi T$ and $i \omega_n = i 2 n \pi T$ are the usual
fermion/boson odd/even Matsubara frequencies ($l$, $n$ integers),
$\xi_{\bf k} = k^2 /(2 m) - \mu$ with $\mu$ the chemical potential and
$m$ the bare band mass, and $\epsilon({\bf k}, i \omega_n)$ is the
dynamical dielectric function, given by the infinite sum of the
polarization bubble diagrams:
\begin{equation}
  \label{eq:epsilon}
  \epsilon({\bf k}, i \omega_n) = 1 - v_q \Pi({\bf k}, i \omega_n),
\end{equation}
with $\Pi({\bf k}, i \omega_n)$ the 3D hole polarizability:
\begin{equation}
  \label{eq:Pi}
  \Pi({\bf k}, i \omega_n) = g \int \frac{d^2 q}{(2 \pi)^2}
  \frac{n_F(\xi_{\bf q}) - n_F(\xi_{{\bf q} - {\bf k}})}
  {\xi_{\bf q} - \xi_{{\bf q} - {\bf k}} + i \omega_n},
\end{equation}
where $n_F(x) = 1/(e^{x/T} + 1)$ is the Fermi distribution function
and $g$ the hole spin degeneracy factor.

In Ga$_{1-x}$Mn$_x$As the valence band structure is complicated with
spin-split light and heavy holes both being important. Fortunately,
our calculated results for BGR do not depend strongly on the carrier
effective mass, and change little if the band mass is changed between
$0.3 - 0.6 m_e$. We choose, consistent with experiment, $m = 0.5 m_e$
as the average band mass in all of our calculations. The dielectric
constant $\kappa$ in the system is $10.9$. The spin degeneracy $g$
varies from $1$ to $2$ according to the magnetization of the DMS
system. ($g=2(1)$ for the paramagnetic (ferromagnetic) case.)
Spin-orbit coupling effects, which are known to be important for GaAs
valence band, are neglected in our theory -- spin-orbit coupling is
expected to lead to small quantitative corrections of our calculated
BGR.

To calculate the retarded self-energy $\Sigma ({\bf k}, i \nu_l \to
\omega + i 0^+) \equiv \Sigma({\bf k}, \omega)$, we perform a certain
contour distortion to transfer the real frequency integration into
summations over imaginary frequencies using the analytic properties of
the dielectric function. This technique is described in detail in
Ref.~\onlinecite{massT}. To obtain quasihole self-energy with
wavevector ${\bf k}$, we simply put $\omega = \xi_{\bf k}$ instead of
solving the Dyson equation $\omega = \xi_{\bf k} +
\mbox{Re}\Sigma({\bf k}, \omega)$. This is known~\cite{rice} to be the
correct approximation consistent with the Fermi liquid theory, within
the single-loop self-energy approximation of Eq.~(\ref{eq:self}).  The
quasihole self-energy then becomes a function of wavevector only:
\begin{eqnarray}
  \label{eq:Esum}
  \Sigma({\bf k}) =&-& \int \frac{d^3 q}{(2 \pi)^3}
  v_q n_F(\xi_{{\bf q} - {\bf k}}) \nonumber \\
  &-& \int \frac{d^3 q}{(2 \pi)^3} v_q \left[
    \frac{1}{\epsilon(q, \xi_{{\bf q} - {\bf k}} - \xi_{\bf k})} - 1 \right]
  \nonumber \\
  &&~~~~~\cdot \left[n_B(\xi_{{\bf q} - {\bf k}} - \xi_{\bf k})
    +n_F(\xi_{{\bf q} - {\bf k}}) \right] \nonumber \\
  &-& \int \frac{d^3 q}{(2 \pi)^3} T \sum_{\omega_n}
  v_q \left[ \frac{1}{\epsilon(q, i \omega_n)} - 1 \right]
  \nonumber \\
  && ~~~~~~~ \times \frac{1}{i \omega_n - (\xi_{{\bf q} - {\bf k}} -
  \xi_{\bf k})}.
\end{eqnarray}
We directly calculate BGR from Eq.~(\ref{eq:Esum}) for various hole
densities, temperatures and spin polarizations by writing $BGR \equiv
|{\rm Re}~\Sigma(k = 0)|$, the magnitude of the real part of band edge
self-energy.

\section{Results}
\label{sec:results}

\begin{figure}[htbp]
\centering \includegraphics[width=3in]{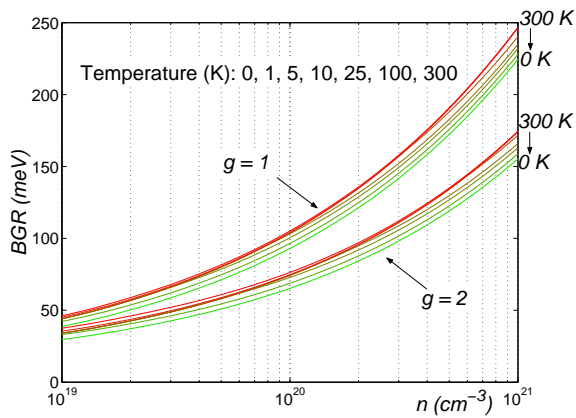}
  \caption{(Color online.) Calculated band-gap renormalization
    (narrowing) as a function of hole-density at different
    temperatures in a ${\rm Ga}_{1-x}{\rm Mn}_x{\rm As}$ system with
    spin degeneracy $g = 1$ and $2$. The upper (lower) 7 curves
    correspond to $g=1 (2)$ case. For each $g$ value, results for 7
    different temperatures are shown: $0K, 1K, 5K, 10K, 25K, 100K,
    300K$. Note that BGR results for $T= 100K$ and $300K$ are almost
    identical, indicating a saturation temperature effect. $E_F =
    (n/g)^{2/3} \times 249.41 meV$ where $n$ is in the unit of
    $10^{20} cm^{-3}$.}
  \label{fig:BGR}
\end{figure}

In Fig.~\ref{fig:BGR} we present the calculated band-gap
renormalization as a function of hole-density $n$ at different
temperatures. We note from Fig.~\ref{fig:BGR} that the BGR reduces the
band gap approximately (but not precisely) as an $n^{1/3}$ functional
dependence on hole density, and the effect can be as large as a few
hundred meVs.  Although BGR itself increases in magnitude with hole
density, we note that the scaled BGR, i.e. $BGR/E_F$, decreases in
magnitude with increasing hole density since $E_F \sim n^{2/3}$.  This
is consistent with the fact that the relative importance of
interaction effects goes down in a quantum Coulomb system with
increasing density. Fig.~\ref{fig:BGR} shows that an increasing
temperature also enhances BGR, and this enhancement has a tendency to
saturate at high temperatures. The increasing BGR with temperature
arises from the weakening of screening with increasing temperature.
The temperature dependence we obtain is moderate but observable. Even
though the percentage correction to BGR due to finite temperature is
larger at lower densities, the absolute finite temperature correction
is approximately the same for all densities within the range $10^{19}
- 10^{21} cm^{-3}$.

\section{Discussion}
\label{sec:diss}

One of the important features of Fig.~\ref{fig:BGR} is that the BGR
depends strongly on the magnetic properties of Ga$_{1-x}$Mn$_x$As with
the ferromagnetic spin-polarized hole $g=1$ situation having larger,
by a factor of $1.5$ to $2$ BGR, than the corresponding
spin-unpolarized ($g=2$) paramagnetic case. The strong spin degeneracy
dependence of BGR apparent in Fig.~\ref{fig:BGR} is understandable on
the basis of the fully spin-polarized system having weaker screening
since the density of states is lower in magnitude in the polarized
system (i.e. $g=1$ is the spin-polarized system versus $g=2$ in the
unpolarized system in Eq.~(\ref{eq:Pi})). Also, BGR depends indirectly
on the Fermi energy, and since the polarized system has a higher $E_F$
than the unpolarized state at the same density, the BGR is higher in
the fully spin-polarized ferromagnetic system. Such a strong
dependence of BGR on the hole spin-polarization (by $50 - 100\%$)
should be reasonably easy to detect experimentally by measuring BGR in
Ga$_{1-x}$Mn$_x$As well below and well above the Curie temperature.

We have also carried out calculations for the quasiparticle effective
mass and spin susceptibility (or equivalently the Landau $g$-factor)
renormalization for the Ga$_{1-x}$Mn$_x$As valence band holes induced
by the hole-hole many-body Coulomb interaction. Our results (not
shown) indicate that many-body hole effective mass and spin
susceptibility renormalizations are rather small ($10-20\%$) at the
hole densities of interest in Ga$_{1-x}$Mn$_x$As. This rather small
quasiparticle Fermi liquid renormalization in Ga$_{1-x}$Mn$_x$As is
consistent with the weakly interacting high-density nature of
Ga$_{1-x}$Mn$_x$As. We have also calculated the hole self-energy
correction $\Sigma({\bf k})$ in Eq.~(\ref{eq:Esum}) as a function of
wavevector to check for the nonparabolicity introduced by hole-hole
interaction. We find that ${\rm Re}~\Sigma(k_F)$ to be slightly ($\sim
10 - 20\%$) larger in magnitude than ${\rm Re}~\Sigma(k = 0)$ at the
hole densities of interest, indicating the correlation-induced
many-body modification of the hole energy dispersion to be rather
small. We have also calculated the imaginary part of the hole
self-energy, ${\rm Im}~\Sigma(k=0)$, to ensure that the quasiparticle
picture does not completely break down at the band edge ($k=0$). We
find $|{\rm Im}~\Sigma(k=0)|/E_F$ to be in the $0.2$ to $0.05$ range
for $n = 10^{19} - 10^{21} cm^{-3}$ hole density range of interest,
indicating that the quasiparticle band description remains
experimentally valid in Ga$_{1-x}$Mn$_x$As in the density range of our
interest.

Now we discuss the effect of impurity scattering on the
Ga$_{1-x}$Mn$_x$As optical properties, which is likely to complicate
the experimental observation of the many-body BGR correction predicted
in this paper, particularly since Ga$_{1-x}$Mn$_x$As samples, even in
the optimally ($x \sim 0.05$) doped metallic system, tend to have
rather large resistivity indicating strong impurity scattering.
Impurity scattering destroys momentum conservation and consequently
may strongly affect inter-band optical absorption experiments which
depend on wavevector conservation. In particular, strong impurity
scattering would lead to two distinct effects on the optical
absorption in heavily doped Ga$_{1-x}$Mn$_x$As. First, there will be
an upward shift in the valence band edge (i.e. a band gap narrowing)
arising from the real part of the self-energy $\Sigma_{\rm i}$
associated with the hole-impurity interaction. This impurity-induced
band-gap narrowing effect has the same practical effect as BGR on the
experimentally measured optical absorption gap, and the net band gap
narrowing will be a sum of the hole-hole self-energy (i.e. BGR) and
the hole-impurity self energy. The hole-impurity self-energy is
therefore significant for the optical absorption experiments of
interest to us. On the other hand the imaginary part of the
hole-impurity self-energy, ${\rm Im}~\Sigma_{\rm i}$, leads to a
broadening of the momentum eigenstates and is therefore a measure of
the level broadening in the optical absorption spectra. This impurity
induced level broadening is, therefore, also an important
consideration for estimating BGR from the optical absorption spectrum
since this will control the broadening of the absorption spectra.

\begin{figure}[htbp]
\centering \includegraphics[width=3in]{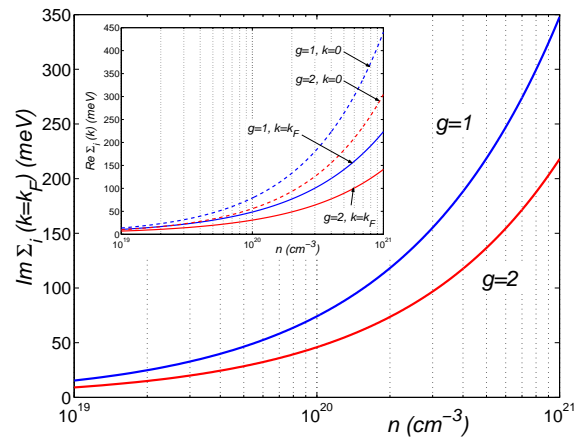}
  \caption{(Color online.) The contribution of impurity scattering 
    to the real and imaginary quasihole self-energy as a function of
    hole-density at $k = 0, k_F$ and $g = 1, 2$ in a ${\rm
      Ga}_{1-x}{\rm Mn}_x{\rm As}$ system. We assume $n_{\rm i} = n$.}
  \label{fig:I}
\end{figure}

We have obtained the impurity scattering effect on the hole states of
Ga$_{1-x}$Mn$_x$As by calculating the hole-impurity self-energy
$\Sigma_{\rm i} ({\bf k})$ in the self-consistent single-site
approximation shown in Fig.~\ref{fig:feynman}(b). The self-consistent
single-site approximation (Fig.~\ref{fig:feynman}(b)), which is a
substantial improvement on the leading-oder Born approximation, should
be qualitatively valid in the metallic regime of Ga$_{1-x}$Mn$_x$As
(as long as the carrier are indeed valence band hole states). The
integral equation represented by the self-consistent hole-impurity
scattering diagrams of Fig.~\ref{fig:feynman}(b) can be exactly solved
for all ${\bf k}$ in the strong impurity screening limit (assuming the
impurities to be random charge center of density $n_{\rm i}$ in
Ga$_{1-x}$Mn$_x$As), and the result for $k=k_F$ and $k=0$ are
\begin{eqnarray}
\label{eq:SigmaI}
\Sigma_{\rm i}(0) &=& V_0 n_{\rm i} (1 + \alpha \sqrt{r_s})^{-1};
\nonumber \\
\Sigma_{\rm i}(k_F) &=& V_0 n_{\rm i} [1 + \alpha \sqrt{r_s} f(\beta
\sqrt{r_s}) + i \gamma],
\end{eqnarray}
where 
\begin{equation}
\label{eq:fx}
f(x) = {1 \over 2} + ({x \over 8} - {1 \over 2 x}) \ln \left| {x + 2
  \over x - 2} \right|,
\end{equation}
and $\alpha = (6g/\pi)^{1/3}$; $\beta = 2^{2/3} 3^{-1/3} \pi^{-2/3}
g^{2/3}$; $\gamma = \pi/2$; $r_s = (9 \pi / 2 g)(a_B k_F)^{-1}$ with
$a_B$ the effective hole Bohr radius in GaAs; $V_0 = 4 \pi e^2/\kappa
q_{TF}^2$ where $q_{TF}$ is the GaAs hole Thomas-Fermi screening
wavevector (and $\kappa$ the GaAs lattice dielectric constant). In
Fig.~\ref{fig:I} we show our calculated real and imaginary parts of
the hole-impurity self-energy assuming an optimal Ga$_{1-x}$Mn$_x$As
metallic system, of $n_{\rm i} \equiv n$. It is important to point out
that our calculated ${\rm Im}~\Sigma$ in Fig.~\ref{fig:I} implies a
level broadening which is consistent with the measured resistivity of
Ga$_{1-x}$Mn$_x$As, providing support for our theoretical
approximation. (The actual $\Sigma_{\rm i}$ may be somewhat smaller
than that given in Fig.~\ref{fig:I} due to impurity clustering effect
ignored in our theory.)  We note that the impurity induced level
broadening, while being somewhat less in magnitude, is of the same
order as the BGR, which may complicate the interpretation of the
optical absorption data. (The net band gap shrinkage is given by the
sum of BGR (Fig.~\ref{fig:BGR}) and the real part of impurity-hole
self-energy given in Fig.~\ref{fig:I}.) But, we believe that it should
still be possible to analyze the optical absorption data to check
whether the density dependent optical absorption spectra are
consistent with a density, temperature, and spin polarization
dependent BGR predicted in our Fig.~\ref{fig:BGR}. Especially, the
temperature dependence of the impurity scattering effect is very
small, and this enables us to identity whether the band shrinking is
indeed a many-body BGR effect instead of impurity scattering by
examining whether this gap shrinking (if observed) possesses an
obvious temperature dependence.

If such a many-body BGR is observed in the experimental data, then
that would be strong evidence supporting a valence band RKKY-Zener
mechanism for DMS ferromagnetism in Ga$_{1-x}$Mn$_x$As.  This is
because if the holes are located in the impurity band, the BGR effect
should be very small because of the large band mass associated with
the impurity band. (Note the impurity band BGR effect just mentioned
should not be confused with impurity scattering effect mentioned in
the last paragraph.) We note that in addition to optical data, STM
measurements~\cite{STM} may also be helpful in the observation of BGR.
We emphasize that because of the large defect and impurity content in
GaMnAs, invariably present~\cite{STM} in the low-temperature MBE
growth of DMS materials, the observation of BGR will be complicated,
but our calculated density, temperature, and magnetization dependence
results should enable such a BGR observation of it is present.  We
also note that at finite temperature electron-phonon
interaction~\cite{cardona} would also contribute to the BGR, but the
phonon effect is smaller in magnitude than the exchange-correlation
correction in the high hole density GaMnAs of interest to us. Also,
the phonon correction does not exhibit a strong density or
magnetization dependence.

In summary, we have developed a theory for hole-hole (and
hole-impurity) free carrier interaction induced many-body effects on
the optical absorption spectra of Ga$_{1-x}$Mn$_x$As, finding (large)
density and magnetization dependent and (moderate) temperature
dependent many-body band gap renormalization corrections, which should
be observable experimentally provided the holes in Ga$_{1-x}$Mn$_x$As
indeed reside in the GaAs valence band, and {\em not} in the impurity
band.

This work is supported by ONR and ARDA.


\bibliography{BGR}

\begin{thebibliography}{20}
\expandafter\ifx\csname natexlab\endcsname\relax\def\natexlab#1{#1}\fi
\expandafter\ifx\csname bibnamefont\endcsname\relax
  \def\bibnamefont#1{#1}\fi
\expandafter\ifx\csname bibfnamefont\endcsname\relax
  \def\bibfnamefont#1{#1}\fi
\expandafter\ifx\csname citenamefont\endcsname\relax
  \def\citenamefont#1{#1}\fi
\expandafter\ifx\csname url\endcsname\relax
  \def\url#1{\texttt{#1}}\fi
\expandafter\ifx\csname urlprefix\endcsname\relax\def\urlprefix{URL }\fi
\providecommand{\bibinfo}[2]{#2}
\providecommand{\eprint}[2][]{\url{#2}}

\bibitem[{\citenamefont{{Das Sarma} et~al.}()\citenamefont{{Das Sarma}, Hwang,
  and Kaminski}}]{DMS}
\bibinfo{author}{\bibfnamefont{S.}~\bibnamefont{{Das Sarma}}},
  \bibinfo{author}{\bibfnamefont{E.~H.} \bibnamefont{Hwang}}, \bibnamefont{and}
  \bibinfo{author}{\bibfnamefont{A.}~\bibnamefont{Kaminski}},
  \bibinfo{howpublished}{Solid State Commun. \textbf{127}, 99 (2003); R. N.
  Bhatt, M. Berciu, M. P. Kennett and X. Wan, J. of Supercon. \textbf{15}, 71
  (2002); C. Timm, J. Phys. Condens. Mat. \textbf{16}, R1865 2003; B. Lee, T.
  Jungwirth, A. H. MacDonald, Sem. Sci. Tech. \textbf{17}, 393 (2002)}.

\bibitem[{\citenamefont{Chattopadhyay et~al.}(2001)\citenamefont{Chattopadhyay,
  {Das Sarma}, and Millis}}]{chattopadhyay}
\bibinfo{author}{\bibfnamefont{A.}~\bibnamefont{Chattopadhyay}},
  \bibinfo{author}{\bibfnamefont{S.}~\bibnamefont{{Das Sarma}}},
  \bibnamefont{and} \bibinfo{author}{\bibfnamefont{A.~J.}
  \bibnamefont{Millis}}, \bibinfo{journal}{\prl} \textbf{\bibinfo{volume}{87}},
  \bibinfo{pages}{227202} (\bibinfo{year}{2001}).

\bibitem[{\citenamefont{Dietl}(2002)}]{dietl}
\bibinfo{author}{\bibfnamefont{T.}~\bibnamefont{Dietl}},
  \bibinfo{journal}{Semicond. Sci. Technol.} \textbf{\bibinfo{volume}{17}},
  \bibinfo{pages}{377} (\bibinfo{year}{2002}).

\bibitem[{\citenamefont{Berciu and Bhatt}(2001)}]{berciu}
\bibinfo{author}{\bibfnamefont{M.}~\bibnamefont{Berciu}} \bibnamefont{and}
  \bibinfo{author}{\bibfnamefont{R.~N.} \bibnamefont{Bhatt}},
  \bibinfo{journal}{\prl} \textbf{\bibinfo{volume}{87}},
  \bibinfo{pages}{107203} (\bibinfo{year}{2001}).

\bibitem[{\citenamefont{Dietl et~al.}(1997)\citenamefont{Dietl, Haury, and
  {Merle d'Aubigne}}}]{dietl2}
\bibinfo{author}{\bibfnamefont{T.}~\bibnamefont{Dietl}},
  \bibinfo{author}{\bibfnamefont{A.}~\bibnamefont{Haury}}, \bibnamefont{and}
  \bibinfo{author}{\bibfnamefont{Y.}~\bibnamefont{{Merle d'Aubigne}}},
  \bibinfo{journal}{\prb} \textbf{\bibinfo{volume}{55}}, \bibinfo{pages}{R3347}
  (\bibinfo{year}{1997}).

\bibitem[{\citenamefont{Akai}(1998)}]{akai}
\bibinfo{author}{\bibfnamefont{H.}~\bibnamefont{Akai}}, \bibinfo{journal}{\prl}
  \textbf{\bibinfo{volume}{81}}, \bibinfo{pages}{3002} (\bibinfo{year}{1998}).

\bibitem[{\citenamefont{Sanvito et~al.}(2002)\citenamefont{Sanvito, G, and
  Hill}}]{sanvito}
\bibinfo{author}{\bibfnamefont{S.}~\bibnamefont{Sanvito}},
  \bibinfo{author}{\bibfnamefont{G.~T.} \bibnamefont{G}}, \bibnamefont{and}
  \bibinfo{author}{\bibfnamefont{N.~A.} \bibnamefont{Hill}},
  \bibinfo{journal}{J. of Supercond.} \textbf{\bibinfo{volume}{15}},
  \bibinfo{pages}{85} (\bibinfo{year}{2002}).

\bibitem[{\citenamefont{{Das Sarma} et~al.}(2003)\citenamefont{{Das Sarma},
  Hwang, and Kaminski}}]{dassarma2}
\bibinfo{author}{\bibfnamefont{S.}~\bibnamefont{{Das Sarma}}},
  \bibinfo{author}{\bibfnamefont{E.~H.} \bibnamefont{Hwang}}, \bibnamefont{and}
  \bibinfo{author}{\bibfnamefont{A.}~\bibnamefont{Kaminski}},
  \bibinfo{journal}{\prb} \textbf{\bibinfo{volume}{67}},
  \bibinfo{pages}{155201} (\bibinfo{year}{2003}).

\bibitem[{\citenamefont{Hirakawa et~al.}()\citenamefont{Hirakawa, Katsumoto,
  Hayashi, Hashimoto, and Iye}}]{hirakawa}
\bibinfo{author}{\bibfnamefont{K.}~\bibnamefont{Hirakawa}},
  \bibinfo{author}{\bibfnamefont{S.}~\bibnamefont{Katsumoto}},
  \bibinfo{author}{\bibfnamefont{T.}~\bibnamefont{Hayashi}},
  \bibinfo{author}{\bibfnamefont{Y.}~\bibnamefont{Hashimoto}},
  \bibnamefont{and} \bibinfo{author}{\bibfnamefont{Y.}~\bibnamefont{Iye}},
  \bibinfo{howpublished}{\prb \textbf{65}, 193312 (2002); E. J. Singley, R.
  Kawakami, D. D. Awschalom, and D. N. Basov, \prl \textbf{89}, 097203 (2002)}.

\bibitem[{\citenamefont{Hwang et~al.}(2002)\citenamefont{Hwang, Millis, and
  Sarma}}]{hwang}
\bibinfo{author}{\bibfnamefont{E.~H.} \bibnamefont{Hwang}},
  \bibinfo{author}{\bibfnamefont{A.~J.} \bibnamefont{Millis}},
  \bibnamefont{and} \bibinfo{author}{\bibfnamefont{S.~D.} \bibnamefont{Sarma}},
  \bibinfo{journal}{\prb} \textbf{\bibinfo{volume}{65}},
  \bibinfo{pages}{233206} (\bibinfo{year}{2002}).

\bibitem[{\citenamefont{Yang et~al.}(2003)\citenamefont{Yang, Sinova,
  Jungwirth, Shim, and MacDonald}}]{yang}
\bibinfo{author}{\bibfnamefont{S.-R.~E.} \bibnamefont{Yang}},
  \bibinfo{author}{\bibfnamefont{J.}~\bibnamefont{Sinova}},
  \bibinfo{author}{\bibfnamefont{T.}~\bibnamefont{Jungwirth}},
  \bibinfo{author}{\bibfnamefont{Y.~P.} \bibnamefont{Shim}}, \bibnamefont{and}
  \bibinfo{author}{\bibfnamefont{A.~H.} \bibnamefont{MacDonald}},
  \bibinfo{journal}{\prb} \textbf{\bibinfo{volume}{67}},
  \bibinfo{pages}{045205} (\bibinfo{year}{2003}).

\bibitem[{\citenamefont{Singley et~al.}(2003)\citenamefont{Singley, Burch,
  Kawakami, Stephens, Awschalom, and Basov}}]{singley}
\bibinfo{author}{\bibfnamefont{E.~J.} \bibnamefont{Singley}},
  \bibinfo{author}{\bibfnamefont{K.~S.} \bibnamefont{Burch}},
  \bibinfo{author}{\bibfnamefont{R.}~\bibnamefont{Kawakami}},
  \bibinfo{author}{\bibfnamefont{J.}~\bibnamefont{Stephens}},
  \bibinfo{author}{\bibfnamefont{D.~D.} \bibnamefont{Awschalom}},
  \bibnamefont{and} \bibinfo{author}{\bibfnamefont{D.~N.} \bibnamefont{Basov}},
  \bibinfo{journal}{\prb} \textbf{\bibinfo{volume}{68}},
  \bibinfo{pages}{165204} (\bibinfo{year}{2003}).

\bibitem[{\citenamefont{{Das Sarma} et~al.}(1990)\citenamefont{{Das Sarma},
  Jalabert, and Yang}}]{dassarma3}
\bibinfo{author}{\bibfnamefont{S.}~\bibnamefont{{Das Sarma}}},
  \bibinfo{author}{\bibfnamefont{R.}~\bibnamefont{Jalabert}}, \bibnamefont{and}
  \bibinfo{author}{\bibfnamefont{S.-R.~E.} \bibnamefont{Yang}},
  \bibinfo{journal}{\prb} \textbf{\bibinfo{volume}{41}}, \bibinfo{pages}{8288}
  (\bibinfo{year}{1990}).

\bibitem[{\citenamefont{Tr\text{\"{a}}nkle
  et~al.}()\citenamefont{Tr\text{\"{a}}nkle, Leier, Forchel, Haug, Ell, and
  Weimann}}]{trankle}
\bibinfo{author}{\bibfnamefont{G.}~\bibnamefont{Tr\text{\"{a}}nkle}},
  \bibinfo{author}{\bibfnamefont{H.}~\bibnamefont{Leier}},
  \bibinfo{author}{\bibfnamefont{A.}~\bibnamefont{Forchel}},
  \bibinfo{author}{\bibfnamefont{H.}~\bibnamefont{Haug}},
  \bibinfo{author}{\bibfnamefont{C.}~\bibnamefont{Ell}}, \bibnamefont{and}
  \bibinfo{author}{\bibfnamefont{G.}~\bibnamefont{Weimann}},
  \bibinfo{howpublished}{\prl \textbf{58}, 419 (1987); G. Tr\text{\"{a}}nkle,
  E. Lach, A. Forchel, F. Scholz, C. Ell, H. Haug, G. Weimann, G. Griffiths, H.
  Kroemer, and S. Subbanna, \prb \textbf{36}, 6712 (1987); G. Bongiovanni and
  J. L. Staehli, \prb \textbf{39}, 8359 (1989)}.

\bibitem[{\citenamefont{Hedin}(1965)}]{hedin}
\bibinfo{author}{\bibfnamefont{L.}~\bibnamefont{Hedin}},
  \bibinfo{journal}{Phys. Rev.} \textbf{\bibinfo{volume}{139}},
  \bibinfo{pages}{A796} (\bibinfo{year}{1965}).

\bibitem[{\citenamefont{Tiago et~al.}(2004)\citenamefont{Tiago, Ismail-Beigi,
  and Louie}}]{louie}
\bibinfo{author}{\bibfnamefont{M.~L.} \bibnamefont{Tiago}},
  \bibinfo{author}{\bibfnamefont{S.}~\bibnamefont{Ismail-Beigi}},
  \bibnamefont{and} \bibinfo{author}{\bibfnamefont{S.~G.} \bibnamefont{Louie}},
  \bibinfo{journal}{\prb} \textbf{\bibinfo{volume}{69}},
  \bibinfo{pages}{125212} (\bibinfo{year}{2004}).

\bibitem[{\citenamefont{Sarma et~al.}()\citenamefont{Sarma, Galitski, and
  Zhang}}]{massT}
\bibinfo{author}{\bibfnamefont{S.~D.} \bibnamefont{Sarma}},
  \bibinfo{author}{\bibfnamefont{V.~M.} \bibnamefont{Galitski}},
  \bibnamefont{and} \bibinfo{author}{\bibfnamefont{Y.}~\bibnamefont{Zhang}},
  \bibinfo{howpublished}{\prb \textbf{69}, 125334 (2004)); Y. Zhang and S. Das
  Sarma, \prb \textbf{70}, 035104 (2004)}.

\bibitem[{\citenamefont{Zhang and {Das Sarma}}()}]{rice}
\bibinfo{author}{\bibfnamefont{Y.}~\bibnamefont{Zhang}} \bibnamefont{and}
  \bibinfo{author}{\bibfnamefont{S.}~\bibnamefont{{Das Sarma}}},
  \bibinfo{howpublished}{\prb \textbf{71}, 045322 (2005); T. M. Rice, Ann.
  Phys. (N. Y.) \textbf{31}, 100 (1965)}.

\bibitem[{\citenamefont{Feenstra et~al.}()\citenamefont{Feenstra, Woodall, and
  Pettit}}]{STM}
\bibinfo{author}{\bibfnamefont{R.~M.} \bibnamefont{Feenstra}},
  \bibinfo{author}{\bibfnamefont{J.~M.} \bibnamefont{Woodall}},
  \bibnamefont{and} \bibinfo{author}{\bibfnamefont{G.~D.}
  \bibnamefont{Pettit}}, \bibinfo{howpublished}{\prl \textbf{71}, 1176 (1993);
  C. Timm, F. Sch\text{\"{a}}er, and F. {von Oppen}, \prl \textbf{89}, 137201
  (2002)}.

\bibitem[{\citenamefont{Cardona}(2005)}]{cardona}
\bibinfo{author}{\bibfnamefont{M.}~\bibnamefont{Cardona}},
  \bibinfo{journal}{Sol. St. Commun.} \textbf{\bibinfo{volume}{133}},
  \bibinfo{pages}{3} (\bibinfo{year}{2005}).

\end{thebibliography}
\end{document}